\documentclass[10pt,letterpaper,compsoc,conference]{iiswc25}

\usepackage{cite}
\usepackage{amsmath,amssymb,amsfonts}
\usepackage{algorithmic}
\usepackage{graphicx}
\usepackage[dvipsnames]{xcolor}
\usepackage[final]{microtype}
\usepackage[italic]{mathastext}
\usepackage{libertine}
\usepackage[T1]{fontenc}
\usepackage{textcomp}
\usepackage[varqu,varl]{zi4}
\usepackage[all]{nowidow}
\usepackage[keeplastbox]{flushend}
\usepackage{fancyhdr}

\usepackage{subfigure}
\usepackage{algorithm}
\usepackage[hidelinks]{hyperref}
\usepackage{colortbl}
\usepackage{multirow}
\usepackage{bm}
\usepackage{makecell}
\usepackage{soul}
\usepackage{enumitem}
\usepackage{capt-of}
\usepackage{balance}

\fancypagestyle{plain}{

    \fancyhf{}
    \fancyfoot[C]{\small \textbf{\textit{Accepted paper at IISWC'25}}\\[0.3em]\thepage}
}

\begin{document}


\title{WANify: Gauging and Balancing Runtime WAN Bandwidth for Geo-distributed Data Analytics





\author{\IEEEauthorblockN{Anshuman Das Mohapatra}
\IEEEauthorblockA{University of Nebraska at Omaha \\
Omaha, Nebraska, USA\\
adasmohapatra@unomaha.edu}
\and
\IEEEauthorblockN{Kwangsung Oh}
\IEEEauthorblockA{\textit{University of Nebraska at Omaha} \\
Omaha, Nebraska, USA \\
kwangsungoh@unomaha.edu}
}}

\maketitle
\thispagestyle{plain}

\begin{abstract}
Accurate wide area network (WAN) bandwidth (BW) is essential for geo-distributed data analytics (GDA) systems to make optimal decisions such as data and task placement to improve performance. Existing GDA systems, however, measure WAN BW \textit{statically} and \textit{independently} between data centers (DCs), while data transfer occurs \textit{dynamically} and \textit{simultaneously} among DCs during workload execution. Also, they use a \textit{single connection} WAN BW that cannot capture actual WAN capacities between distant DCs. Such inaccurate WAN BWs yield
sub-optimal decisions, inflating overall query latency and
cost. In this paper, we present \textit{WANify}, a new framework that
precisely and dynamically gauges \textit{achievable} runtime WAN
BW using a machine learning prediction scheme, decision tree-based Random Forest. This helps GDA systems make better decisions yielding reduced latency and costs including WAN BW monitoring costs.
Based on predicted runtime WAN BW,
WANify determines the optimal number of \textit{heterogeneous parallel connections} for data transfer among DCs. This approach improves performance without additional, or even at reduced cost, 
by fully exploiting available WAN capacities.
In addition, WANify considers dynamics like network and workloads, and heterogeneity like skewed data,
heterogeneous compute resources, and a varying number of
DCs while making decisions. The WANify prototype running on state-of-the-art GDA systems is evaluated on AWS
with 8 geo-distributed DCs. Results show that
WANify enhances WAN throughput by balancing between the strongest and weakest WAN links, enabling GDA systems to reduce latency and cost by up to 26\% and 16\% respectively with minimal effort, all while handling dynamics and heterogeneity efficiently. 
\end{abstract}



\section{Introduction} \label{sec:intro}
Many multi-region cloud applications like Netflix, Alibaba, and Tencent 
generate large-scale data in a geo-distributed manner. For example, Netflix alone operates in 233 
regions to serve their geo-distributed users.
\vspace*{\fill} 
\noindent\rule{\columnwidth}{0.4pt} 
{%
\footnotesize
\raggedright
\noindent
\copyright~2025 IEEE. Personal use of this material is permitted.  Permission from IEEE must be obtained for all other uses, in any current or future media, including reprinting/republishing this material for advertising or promotional purposes, creating new collective works, for resale or redistribution to servers or lists, or reuse of any copyrighted component of this work in other works.
}

\newpage
\noindent To comply with regional data protection laws \cite{bergui2021survey, rojszczak2020cloud} and achieve the desired performance \cite{Kimchi}, these applications rely on geo-distributed data analytics (GDA) to gain meaningful and timely knowledge from a highly dispersed 
large volume of data \cite{NetflixDistTraining, Yugong, Tencent}. Consequently, since these results are used to make important decisions affecting revenues and system health, improving their performance is a critical goal.

While de-facto data processing frameworks, like Hadoop \cite{hadoop} and Spark \cite{spark},
have been widely used for analytics jobs, they are known to be ineffective in GDA \cite{cardosa2011exploring}. 
This is because large data must be transferred via wide area network (WAN), one of the most scarce and expensive resources.
That is, WAN offers lower bandwidth (BW)\footnote{We use BW to refer to WAN throughput unless stated otherwise.} than other mediums, e.g., memory, SSD, HDD, and LAN, within a data center (DC) \cite{DelayResistantMostafaei}.
Thus, existing GDA systems \cite{Iridium, Kimchi, Tetrium, Yugong} 
consider heterogeneous BW while making optimal decisions, 
e.g., task and data placement, that minimize data transfer via weak WAN links
to avoid the performance bottlenecks, i.e., heterogeneous BW-aware approaches. 

To identify bottlenecked WAN links, these systems measure BW between 
the virtual machines (VMs) deployed in geo-distributed DCs \textit{statically} and \textit{independently} using host-based network tools, e.g., iPerf \cite{iperf}.
They use these BWs as inputs for making optimal BW-aware decisions at several stages of a GDA query.
Such an approach assumes that runtime BW during query execution will be the same with (or similar to) the statically measured BW.
This assumption, however, could easily become \textit{false} because data transfer occurs \textit{dynamically} 
and \textit{simultaneously} in GDA (especially shuffle stages).
That is, BW among DCs can vary based on the number of egress/ingress connections that are active at runtime. 
Our preliminary experiments confirmed that statically and independently measured BWs do not represent 
runtime BW, as Section \ref{subsec:illEx} will show. 
This implies that existing BW-aware GDA systems \cite{Iridium, Kimchi, Tetrium, Yugong} may have produced sub-optimal solutions 
due to inaccurate BW values, which limits their performance gains. 
While measuring BWs across all DC pairs concurrently could provide
accurate runtime BW, it incurs significant costs (Section \ref{subsec:illEx}). Therefore, gauging runtime BW precisely would be highly desirable for them to make
optimal decisions while avoiding cost bottlenecks.

The gains from these systems could be further sub-optimal because they use a \textit{single connection} to transfer data across DC-pairs, 
which cannot capture available BW between the distant DCs \cite{patthel}.
\begin{figure}[tp]
	\centering
	\includegraphics[width=2.3in]{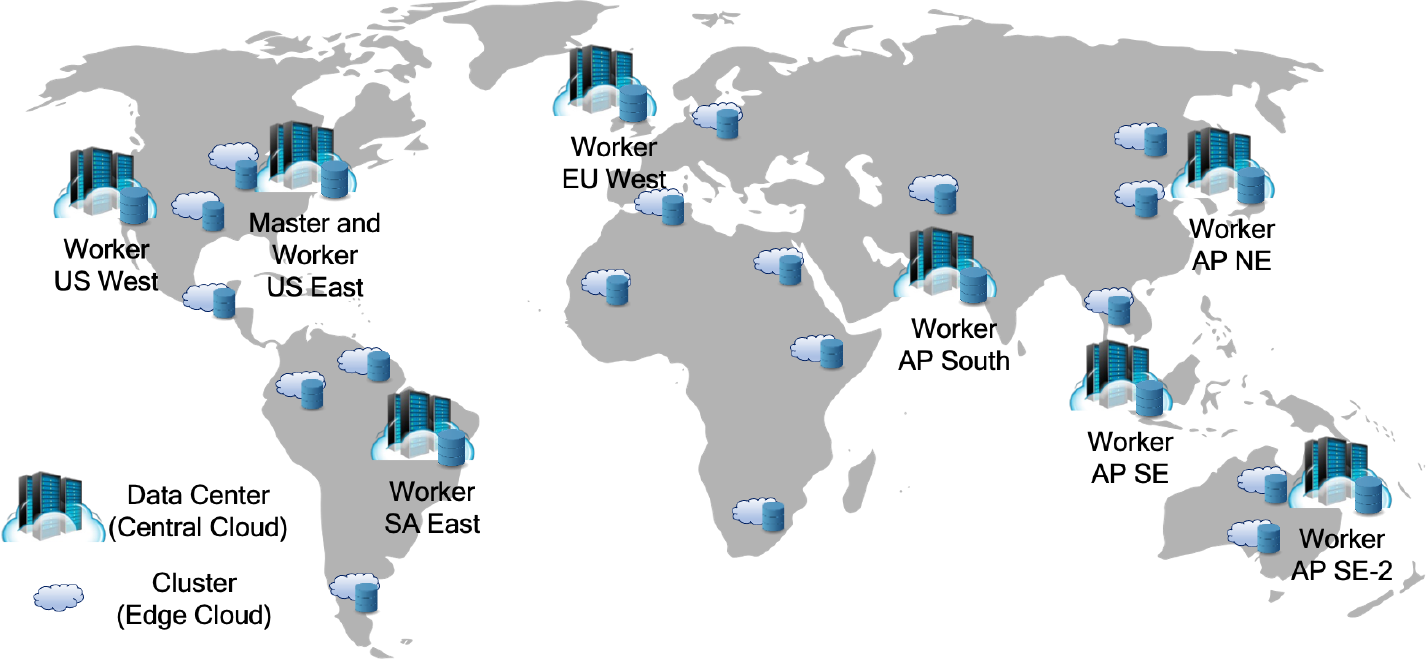}
	\caption{Highly geo-distributed data and cloud resources}
	\label{fig:gda}
\end{figure}
With a single connection, BW between faraway DCs is often notably lesser than the closer ones due to a greater number of network hops.
Given the DCs in Fig. \ref{fig:gda}, for example, 
we observed the weakest BW (121 Mbps) between US East and Asia Pacific (AP) SE and the highest BW (1700 Mbps) between US East and US West using a single connection.
Based on such BWs, GDA systems tried to minimize
transferring data via weak WAN links, e.g., between US East and AP SE.
However, it is known that WAN performance can be improved through parallel data transfers using multiple connections (channels) \cite{ito2006gridftp, skyplane, patthel}. In our experiments, the weakest link between US East and AP SE increased up to 1 Gbps using 9 connections.
Since GDA performance is highly affected  
by weak (bottlenecked) WAN links, improving their performance by 
using multiple connections would reduce query latency and cost.

However, obtaining accurate runtime BWs and fully utilizing available BW are far from trivial.
That is, gauging runtime BW precisely is challenging 
as BW is affected by several metrics such as the number of active peers and channel utilization of each peer.
To improve WAN performance, one may uniformly increase the connections for each DC pair. However, this would incur a \textit{BW-starvation} between distant DCs 
as nearby DCs occupy most of the available network, 
i.e., race condition and network contention.
Thus, a heterogeneous number of connections should be determined optimally.
This approach, however, is exponentially complex, e.g., $10^{240}$ different possibilities 
with 240 peering connections for two workers in each of the 8 different regions (with a maximum of 10 parallel connections between each DC-pair). Lastly, dynamics and heterogeneity in GDA, e.g., fluctuating BWs \cite{FluctuatingBWsIMC}, varying number of DCs (and VMs), and input data skewness, make it impossible to 
determine optimal decisions statically.
To address these challenges, we present a new WAN framework called \textit{WANify}. WANify gauges runtime BW based on real-time snapshots and diverse inputs, e.g., number of DCs and their physical distance, 
which allows GDA systems to make optimal decisions by using accurate BWs. Also, WANify actively determines optimal and heterogeneous number of connections 
across all the peer-to-peer DC links. These determinations rely on the \textit{achievable} predicted BW
in order to 
improve weak WAN links, which would improve overall query latency by fully exploiting the WAN capacities. In addition, WANify uses a distributed agent-based model for handling dynamics and considers many forms of heterogeneity, such as skewed data, heterogeneous compute resources, and a varying number of DCs, while making optimal decisions. 
Note that WANify considers WAN performance of VMs as a user of cloud providers. Thus, it does not change anything below the application layer and because of its lightweight and orthogonal design, it can be easily integrated into any GDA system. 

We implemented WANify prototype on Spark and evaluated it on AWS virtual private cloud (VPC) using state-of-the-art WAN-aware GDA systems, i.e., Tetrium \cite{Tetrium} and Kimchi \cite{Kimchi}, to show how they can benefit from WANify with minimal changes. Further, we examined WANify's generalizability on a recent BW-aware machine learning (ML) workload in GDA, i.e., SAGQ \cite{quantizationTCC}.
Results show that WANify reduces latency and cost by up to $26\%$ and $16\%$, respectively with a $2\times$ increase in the minimum BW of the cluster. Also, the results confirm that WANify helps GDA systems handle dynamics and heterogeneity efficiently.

Overall, the research contributions of this paper are:
\begin{itemize}[noitemsep,topsep=0pt]
	\item The design of WANify, which is the first runtime parallelization-enabled WAN-BW-optimized framework (to the best of our knowledge) for GDA.
    \item Gauging runtime WAN BWs precisely, which enables GDA systems to derive better solutions at lower cost.
    \item The idea of using heterogeneous connections in GDA, which reduces latency by trading off the strong WAN links to improve the weak WAN links of the cluster.
	\item Considering dynamics and heterogeneity, e.g., fluctuating BW, changes to cluster sizes, and the presence of skewed data, to avoid cost-performance bottlenecks. 
	\item Thorough empirical evaluations on AWS using several baselines, demonstrating the efficacy of WANify.
	\item The open-sourced WANify project, along with the collected datasets \cite{WANify_repo}, to aid future WAN-aware systems.
\end{itemize}

\section{System Model and Motivation}\label{sec:background}

\subsection{System Model}\label{subsec:sysModel}
\noindent\textbf{Data Center (DC) Setting:}
We consider GDA systems running in a highly dispersed DC cluster, as shown in Fig. \ref{fig:gda}.
In this environment, BW among DCs is highly heterogeneous and limited.
This is because cloud providers limit network performance based on compute resources' type and size, and BW is further throttled based on their WAN policies.
For example, AWS offers up to 10 Gbps network BW (sum of inbound and outbound) for m5.large virtual machine (VM) instance, but its WAN BW is reduced by half, i.e., up to 5 Gbps.
In addition, the number of network hops vary based on DC regions, which results in heterogeneous BW among DCs.
For instance, 
we observe the maximum BW between US East and US West (1700 Mbps) and the minimum BW between US East and Asia Pacific (AP) SE (121 Mbps) using iPerf in Fig. \ref{fig:gda}.
Thus, we consider WAN to be the major performance bottleneck like other existing GDA systems \cite{Iridium, Kimchi, Tetrium, Yugong} and underscore the importance of utilizing available BW efficiently for improved query performance.
We are also mindful of various types of fluctuating BWs \cite{FluctuatingBWsIMC}, enabling WANify to handle 
diverse private and public networks, including edge-cloud and VPC.

\noindent\textbf{WAN-aware GDA Systems:}
GDA systems process and analyze geo-distributed large-scale data. They may migrate all input data into a centralized single DC and process it using a parallel processing framework 
like Spark. 
While this is simple, it will cause performance bottlenecks because large input data must be sent via WAN \cite{Kimchi}. Moreover, data protection laws hinder the aggregation of (raw) geo-distributed data centrally \cite{bergui2021survey, rojszczak2020cloud}.
More suitably, input data can be processed in-place to minimize data transfer among DCs.
However, large intermediate data still needs to be sent across DCs, i.e., shuffle stages, which 
incurs performance bottlenecks. 
Since job completion time (JCT) depends on the slowest task that requires data transfers via the \textit{weakest} WAN link, recent GDA systems \cite{Iridium, Kimchi, Tetrium, Yugong}, including distributed ML systems \cite{Gaia, quantizationTCC}, have focused on minimizing the usage of weak links, i.e., heterogeneous WAN-aware approach.
To achieve this, they measure BWs statically and independently to identify weak links while making optimal decisions. 
Data transfer, however, occurs dynamically and simultaneously among DCs via all-to-all communication during job execution, making statically measured BW incorrect.
Thus, prior systems would make sub-optimal decisions, causing performance- and cost-bottlenecks.

\noindent\textbf{Parallel Connections for Data Transfer:}
Existing GDA systems \cite{Iridium, Kimchi, Tetrium, Yugong} transfer data among DCs
using a single connection. This is because they are based on 
de-facto distributed data analytics systems, e.g., Spark, that
were designed for a single DC setting where a single connection can fully utilize
available network BW.
A single connection, however, cannot fully exploit \textit{available} WAN throughput 
between distant DCs \cite{patthel}. In WAN, using parallel connections
is a known technique for improving WAN throughput \cite{Globus, skyplane}.
In our example, BW between US East and AP SE increased to 1 Gbps (from 121 Mbps) using 9 connections.
Additionally, existing GDA systems have determined optimal decisions
based on statically measured BW using a single connection, implying that 
they have not fully utilized the available BW between distant DCs.
Since JCT is determined by the slow tasks using the weak WAN links,
improving their performance by using multiple connections would improve the overall performance.

\begin{table}[t]
	\centering
	\caption{Gaps between static and runtime BWs (Mbps)}
	\scalebox{0.83}{
		\begin{tabular}{|c||c|c|c|}
			\hline
			\textbf{Difference Interval} & \textbf{(100, 200]} & \textbf{(200, 250]} & \textbf{$\bm{>}$ 250} \\
			\hline
			\hline
			\textbf{Count} & \textbf{7} & \textcolor{blue}{\textbf{8}} & \textcolor{red}{\textbf{3}} \\
			\hline
	\end{tabular}}
	\label{tab:bwDiffs2}
\end{table}
\subsection{Motivations}\label{subsec:illEx}
\noindent\textbf{Gaps between Statically Measured BW and Runtime BW:}
We ran iPerf experiments on AWS using 8 DCs in Fig. \ref{fig:gda}, which are connected through VPC peering. We measured one DC-pair BW at a time for static-independent BWs as done in existing  GDA systems. For runtime BW observed during GDA query execution, we measured all DC-pair BWs simultaneously.
Table \ref{tab:bwDiffs2} shows the results with 18 \textit{significant differences}
between the two approaches. 
Note that we consider the difference to be significant 
when it is greater than 100 Mbps since that has been used
as a boundary to characterize network performance \cite{kim2019delay, gomez2020performance}, and inter-DC links often support this as a baseline BW \cite{chen2021sdtp}.
The above change in BW distribution is also accompanied by a change in the DC characteristics. For instance, AP SE is the slowest DC from South America (SA) East in the static approach. Based on this, prior works \cite{Tetrium, Iridium, Kimchi} choose to migrate input data out of AP SE to the nearby DCs.
In the runtime approach, however, EU West is the slowest DC from SA East. This implies that prior GDA systems have made sub-optimal decisions with static BW, limiting their performance gains as Section \ref{subsec:singleConImpr} will show. 
Thus, runtime BW is central for efficiently handling the GDA workloads.

\noindent\textbf{Cost-bottleneck for Measuring BW:}
WAN is one of the most expensive resources in GDA. Consequently, measuring BW frequently could be very costly. If $x$ is the average per-instance-second cost for a cluster with $N$ nodes, $y$ is the monitoring duration, and $z$ is the average per-instance-network cost for data exchanged during monitoring, then each monitoring would cost $N \times (x \times y + z)$. Moreover, monitoring should be performed every few minutes due to the highly dynamic network in GDA \cite{Tetrium, FluctuatingBWsIMC}. Thus, if $O$ is the total occurrences of monitoring in a year, then the annual costs for runtime monitoring is expressed by Eq. \ref{eq:monitoringCosts}.
\begin{equation}
\scalebox{0.85}{$
\begin{aligned}
    \bm{O \times N \times (x \times y + z)}
    \label{eq:monitoringCosts}
\end{aligned}
$}
\end{equation}
Such monitoring costs may incur a cost bottleneck based on these parameters, especially $O$, $N$, and $z$, as network or data transfer cost ($z$) is more expensive than compute cost ($x \times y$) in GDA \cite{Kimchi}. 
\begin{table}[t]
\centering
\caption{Accurate prediction saves $\sim96$\% in costs}
\scalebox{0.85}{\begin{tabular}{|c|c|c|c|}
\hline
\multirow{2}{*}{\textbf{\textit{Number of DCs}}} & \multirow{2}{*}{\textbf{Runtime Monitoring}} & \multicolumn{2}{c|}{\textbf{Prediction Model}} \\
\cline{3-4}
& & Training & Predictions \\
\hline
4& $\$703$&$\$35$ &$\$29$\\
\hline
6& $\$1055$&$\$20$ &$\$16$\\
\hline
8& $\$1406$&$\$14$ &$\$11$\\
\hline
\hline
\textbf{Total} &$\$\bm{3164}$ &$\bm{\$69}$ & $\bm{\$56}$\\
\hline
\end{tabular}}
\label{tab:costSavings}
\end{table}
Also, several applications may require monitoring for different cluster sizes ($N$) \cite{multiClusterSizeNeed1}.
Consequently, having a prediction model that uses \textit{real-time} snapshots or short-duration ($1$ second) inputs to accurately gauge stable runtime BWs for various cluster sizes would drastically reduce the monitoring costs by reducing $y$ and $z$. Empirical results on AWS suggest that stable runtime BWs are achieved with at least 20 seconds of monitoring, and they have a positive Pearson Correlation \cite{cohen2009pearson} with 1-second snapshots. Hence, it is possible to greatly reduce these costs by using a prediction model for multiple cluster sizes. Table \ref{tab:costSavings} shows the potential cost savings for BW monitoring on $3$ different cluster sizes, with $1$ VM per DC. 
Here, costs are derived based on the suggestion of Tetrium \cite{Tetrium}, i.e., measuring BW every few (30) minutes using a t3.nano AWS VM. Also, prediction costs consider a training dataset of 1000 samples having both snapshot and stable runtime BWs. Lastly, network cost is based on average BW of 200 Mbps.

\begin{figure}[t!]
    \centering
	\subfigure[Single connection BWs]{\label{subfig:ill1}\includegraphics[width=0.95in]{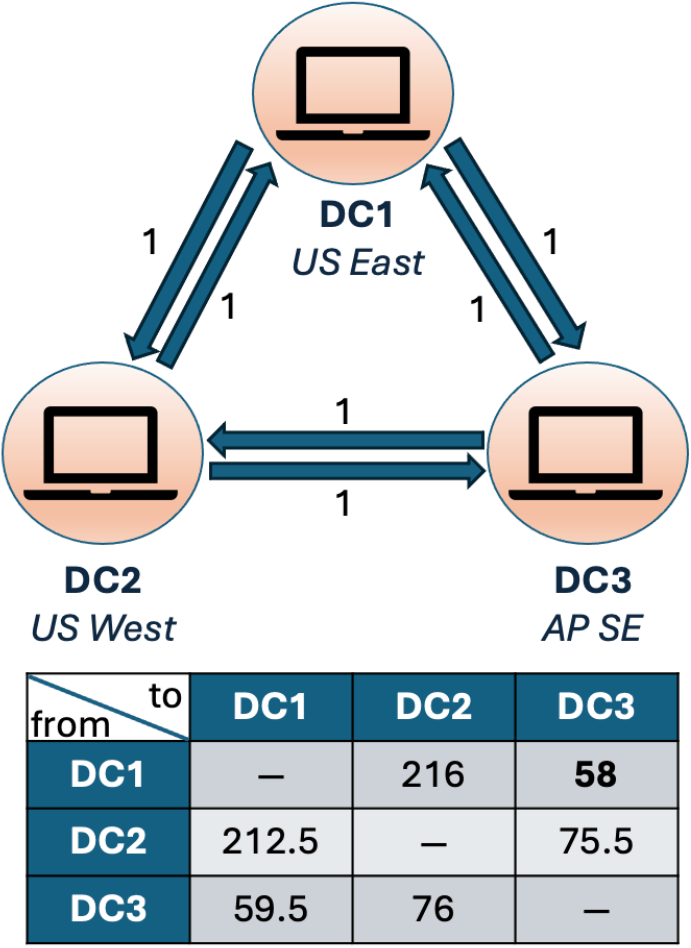}}
        \hfill
	\subfigure[Uniform parallel connections BWs]
	{\label{subfig:ill2}\includegraphics[width=0.95in]{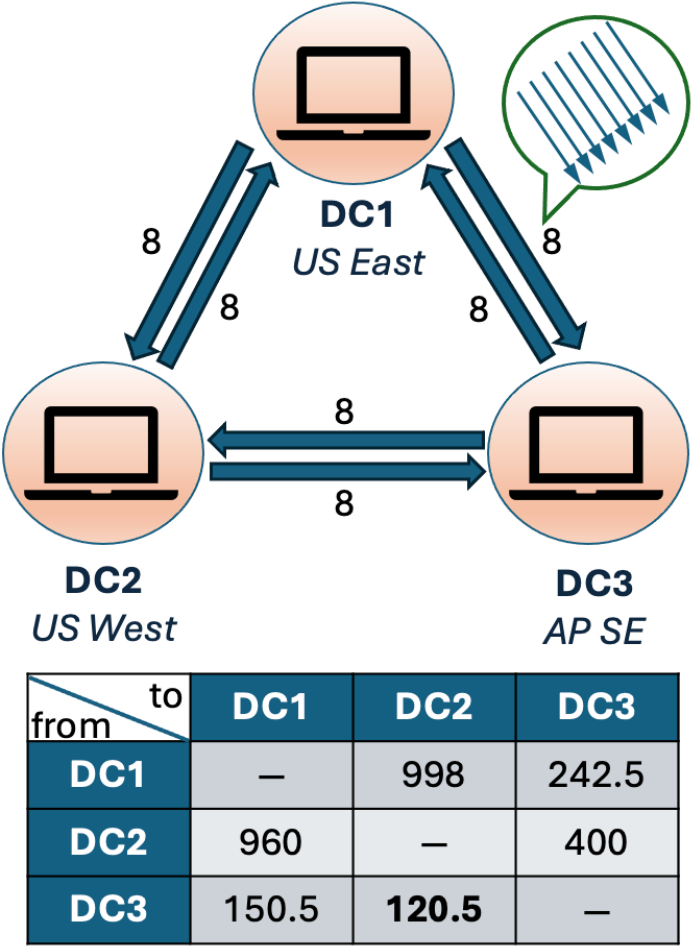}}
	\hfill
	\subfigure[Heterogeneous connections BWs]
	{\label{subfig:ill3}\includegraphics[width=0.95in]{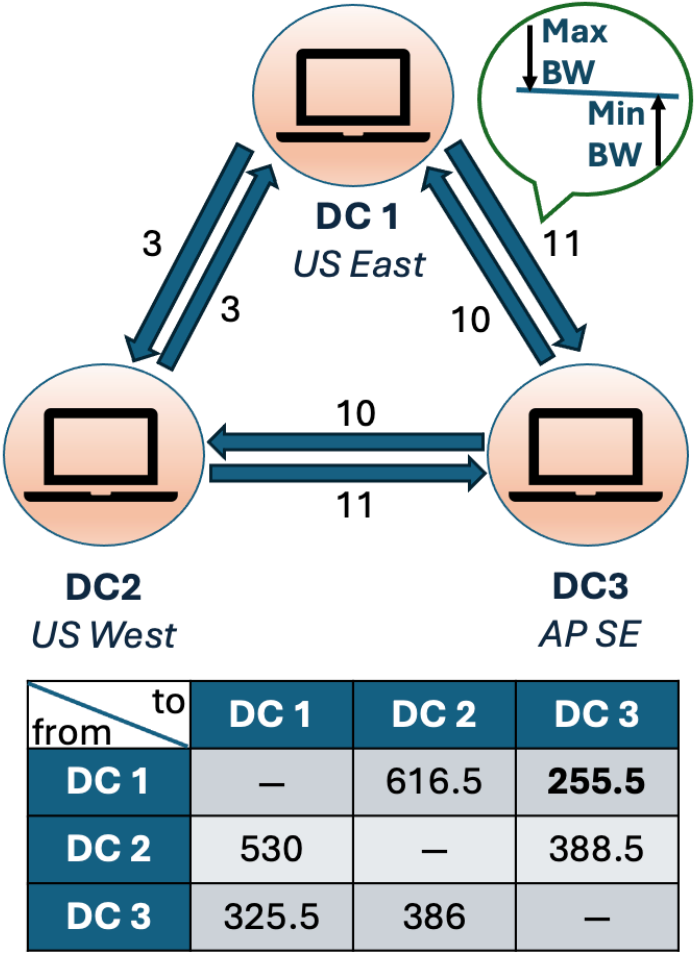}}
	\subfigure[Network overhead]
	{\label{subfig:ill4}\includegraphics[width=1.8in]{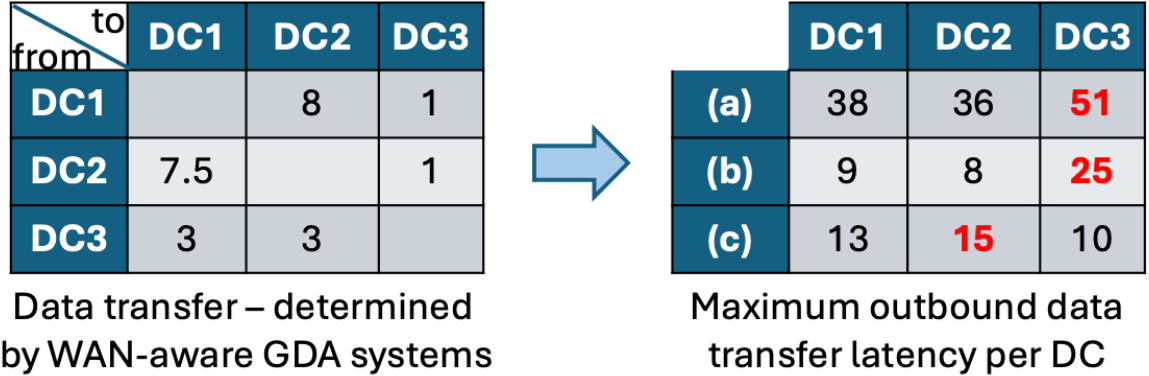}}
	\hfill
	\caption{BWs and network latency for different approaches}
	\label{fig:illEx}
\end{figure}
\noindent\textbf{Heterogeneous connections for Data transfer:}
To check the benefits of parallel heterogeneous data transfer,
we ran more experiments as shown in Fig. \ref{fig:illEx}. 
For runtime BW monitoring (first three figures), each DC has an unlimited-burst \textit{t3.nano} VM and runs iPerf simultaneously to other DCs. This makes a total of 6 unique links with bidirectional concurrent communications among the $3$ DCs. Fig. \ref{subfig:ill1} shows the BWs observed when $1$ connection is used for each link. This results in decent BWs between the nearby DCs, i.e., \textit{DC1} and \textit{DC2}, whereas weak BWs between faraway DCs, i.e., \textit{DC1} and \textit{DC3}, and \textit{DC2} and \textit{DC3}, respectively. Fig. \ref{subfig:ill2} shows the BWs when each link is uniformly increased to $8$ parallel connections. This uniform parallelism has little benefit as nearby DCs occupy most of each other's available network capacity, evidenced by a mere BW of $120.5$ Mbps between \textit{DC3} - \textit{DC2}. Note that increasing link parallelism beyond $8$ resulted in no improvement to the maximum BW (\textit{DC1}$-$\textit{DC2}) because of anticipated network congestion \cite{tcpParallelBeyondLimit1}. 

Alternatively, better approach is shown in Fig. \ref{subfig:ill3}. For simpler comparison with Fig. \ref{subfig:ill2}, it assumes the same total parallel connections ($8 \times 6$),
but heterogeneously distributes them so that the faraway DCs get higher precedence over nearby DCs. Note that although this leads to a reduction in the maximum BW between \textit{DC1} and \textit{DC2}, it improves the weak BW links between other DCs and \textit{DC3}, resulting in a $2.1$ fold increase in the minimum BW from $120.5$ Mbps to $255.5$ Mbps. 
Thus, compared to others, the heterogeneous approach provides a better distribution of BWs that elevate the weak WAN links, reducing the network bottleneck.

To show the benefits of increased minimum BW in a cluster, let us consider $1$ reduce stage in a GDA query, running on the $3-$DC cluster set-up as above. For simplicity, we use data sizes in \textit{Gigabit (Gb)}, since BWs are in Mbps. Also, suppose that a WAN-aware GDA system has found the weak WAN links to \textit{DC3}, and thus, a smaller amount of intermediate data is scheduled for exchange with \textit{DC3} as shown in Fig. \ref{subfig:ill4}. Note that such data is not $0$ at \textit{DC3} since it requires input data migration, which is slow and costly. Next, we compute network latency for various approaches and flag the slowest network time that proves bottleneck for the data transfer operation. 
It is evident that the heterogeneous approach greatly lowers the network time over others, thereby notably lowering the JCT.

Note that the connections in Fig. \ref{subfig:ill3} were found manually for illustrations. 
Thus, swiftly optimizing them at each stage of a GDA query would lower the JCT.
However, as noted in Section \ref{sec:intro}, this task is highly complex.
\subsection{Determining BWs and  Optimal Connections}
In this paper, predicted runtime BWs and optimal heterogeneous connections are derived using decision tree-based approach (Section \ref{subsec:wanPredModel}) and distributed BW optimization (Section \ref{subsec:optimalconnect}). 
Similar to Fig. \ref{subfig:ill3}, these values are represented as $2$ matrices where each cell contains pair-wise BW and the number of connections, respectively. Also, since these values are each structured as a matrix, existing GDA systems on single connection static-independent BWs can take advantage of the proposed mechanisms with minimal effort.

\section{BW Gauging and Optimization} \label{sec:approaches}
We are mindful that accurately determining runtime BW is challenging because of the underlying network dynamics \cite{FluctuatingBWsIMC} and cost (Section \ref{subsec:illEx}). Thus, we use a low-cost prediction model with real-time snapshots to accurately gauge only the current network state for first-hand optimizations, though single connection GDA systems can use this information directly. However, to handle dynamics occurring during the query execution flow, we rely on local fine-tuning as explained later in Section \ref{subsubsec:localOptim}.

\subsection{Offline Runtime BW Prediction Model} \label{subsec:wanPredModel} 
The problem of predicting the runtime BWs between a given number of DCs is a multivariate regression problem with many outliers based on the observed range of BWs. 
Since these outliers can deleteriously affect the statistical regression techniques \cite{osborne2019power}, we avoid them in this work. 
Consequently, we propose a decision tree-based Random Forest (RF) regressor, since they provide desirable results on multivariate regression problems \cite{borchani2015survey} by using ensemble learning to avoid model over-fitting. Another reason for choosing RF is that compared to other ML approaches, such as Support Vector Machines and Decision Trees, RF is known to work well with networked applications \cite{WhyRFModel}. Moreover, RF is computationally less intensive and requires significantly lesser training data than other deep learning techniques \cite{Smartpick}. Note that we initially tried employing Convolutional Neural Network (CNN)  \cite{nakandala2020incremental, medera2009incremental} for this task by using GPUs on AWS, but that did not yield promising results, i.e., it resulted in $\sim85\%$ training accuracy with a higher number of pair-wise BW differences against the test dataset. This is because CNN, a deep learning approach, requires large training data to attain the desired accuracy.

\begin{table}[t]
	\centering
	\caption{Features for Runtime BW Prediction}
	\scalebox{0.85}{\begin{tabular}{|l|l|}
			\hline
			\textbf{Feature} & \textbf{Description} \\
			\hline
			\hline
			$N$ & Number of DCs in the VM-based cluster \\
			$S\_BW_{ij}$ & Real-time snapshot BWs between VMs at DCs i and j \\
			$M_{d}$ & Memory utilization at receiving end \\
			$C_{i}$ & CPU Load at the VM in DC i \\
			$N_{r}$ & Number of retransmissions \\
			$D_{ij}$ & Physical distance (in miles) between VMs at DCs i and j \\
			\hline
	\end{tabular}}
	\label{tab:predictFeatures}
\end{table}

Having identified the 
suitable BW prediction technique,
we explored various features for training the prediction model. Table \ref{tab:predictFeatures} outlines these features, which are derived based on empirical evidence. 
Note that memory utilization ($M_d$) is vital as each connection requires a memory buffer, affecting runtime BW \cite{memoryOnNet}.
Also, physical distance between DCs is a primary feature over hop-count as geo-location has a larger impact on network delay \cite{fei1998measurements}. 
This parameter ($D_{ij}$) is derived using geo-coordinates of VM IPs in the respective DCs. Once datasets containing these features are generated, the RF component is trained to predict runtime BWs ($R\_BW_{ij}$). 
Existing WAN-aware GDA systems can use the predicted BWs as inputs to make optimal decisions. 

\subsection{Determining Optimal Connections}
\label{subsec:optimalconnect}
Given the predicted runtime BWs, we analyze them on the fly and optimize the heterogeneous connections on derived inter-DC relationships (\textit{strong vs weak}), thereby maximizing the achievable BWs.
To handle dynamics in GDA,
this is done using (1) global and (2) local optimizations. While (1) optimizes connections based on the given network state, 
(2) fine-tunes them at runtime for persistent benefits.
\subsubsection{Static Global Optimization}\label{subsubsec:globalOptim}
Algorithm \ref{alg:global_alg} presents the pre-processing step for the global optimization process that consists of a function named INFER\_DC\_RELATIONS (lines 1 - 22). 
It takes the inputs: (1) \textit{bw} - predicted runtime BW computed by the RF and
(2) \textit{D} - the minimum difference between BWs to be considered significant.
These inputs create boundary conditions (lines 3-8) for deriving the physical closeness between a DC-pair, i.e., closeness index (lines 9-22).
For example, given $bw =$ \{1000, 400, 120; 380, 1000, 130; 110, 120, 1000\} and \textit{D} = 30 for three DCs, the algorithm first determines unique BWs: \{110, 120, 130, 380, 400, 1000\} (line 3), followed by filtering using D to generate \{110, 380, 1000\} (lines 4-8). 
Eventually,
closeness values are assigned to each DC-pair using values in $bw$ (lines 9-22).
That is, 1000 is assigned closeness index 1 (highest closeness), \{400, 380\} are assigned closeness index 2, and \{120, 130, 110\} are assigned closeness index 3 (lowest closeness). 

The outputs from Algorithm \ref{alg:global_alg} are used in global optimization, which uses the inferred DC relationships and computes an optimal range of network configurations (such as minimum and maximum values for number of connections and achievable BWs, respectively) in a greedy manner. 
\renewcommand{\algorithmiccomment}[1]{//#1}
\begin{algorithm}[h]
	\caption{Inferring Data Center Relationships}\label{alg:global_alg}
        \scriptsize
	\begin{algorithmic}[1]
		\STATE  \textbf{function} $INFER\_DC\_RELATIONS (bw,D)$
		\STATE \COMMENT $bw$ has runtime BWs and $D$ is min. difference to infer DC relationships
		\STATE $bw_u = sort(set(bw)),\text{ } N = size(bw),\text{ } N_u = rowSize(bw_u)$ \COMMENT{$bw_u$ is a set of BWs}
		\FOR[Reverse traversal for correct deletion of elements]{i = $N_u$ to $2$}
		\IF{$bw_u[i] - bw_u[i-1] < D$}
		\STATE $remove\ bw_u[i]$
		\ENDIF
		\ENDFOR
		\STATE Initialize $DC_{rel}$ to all ones
		\FOR{i = $1$ to $N/2$}
		\FOR{j = $1$ to $N/2$}
		\STATE $k = binarySearch(bw[i][j], bw_u)$ \COMMENT{returns (1-based) index if found or interval otherwise}
		\IF{match found}
		\STATE $DC_{rel_{ij}} = len(bw_u) - k + 1$
		\ELSE
		\STATE $m_1 = k_1\ and\ m_2 = k_2$
		\STATE $closr_{val} = m_1\ or\ m_2$\COMMENT{based on distance to $bw[i][j]$}
		\STATE $DC_{rel_{ij}} = len(bw_u) - closr_{val} + 1$
		\ENDIF
		\ENDFOR
		\ENDFOR
		\STATE return $DC_{rel}$
	\end{algorithmic}
\end{algorithm}
To achieve this, it first computes the sum of $DC_{rel}$ (received from Algorithm \ref{alg:global_alg}), i.e. \textbf{($sum_{all}$}, and row-wise maximums, i.e. \textbf{$max_{r_i}$}, as shown in Eq. \ref{eq:forall_preCondition}. Subsequently, it uses these values in Eq. \ref{eq:forall_Eq} with $minCandidate$ and $maxCons$, and favors DC pairs with a higher closeness index, implying that distant DCs with low BWs get a higher number of connections from a reference DC that has limited number of total parallel connections ($M$). 
This is because the maximum parallel connections from a VM in a given DC is limited, and increasing connections beyond this optimal threshold causes performance degradation, as shown in Section \ref{subsec:illEx}.
Note that $sum_{all}$ skips closeness index $1$ (all diagonal elements in $DC_{rel}$) 
since a single connection within a single DC can fully utilize the available BWs as explained in Section \ref{subsec:sysModel}.
Using the values from the example in Algorithm \ref{alg:global_alg} above and $M = 8$ yields $minCons$ with all ones and $maxCons$ as \{3, 6, 8; 6, 3, 8; 8, 8, 3\}.
Also, empirical results confirm that runtime BW grows linearly with the connections, and thus, in Eq. \ref{eq:forall_Eq}, we use the product of predicted BW and determined connections as the achievable BW of a DC pair.
Further, this equation considers weights ($w_s$) for skewed input data and refactoring-vector ($r_{vec}$) for heterogeneous cloud providers, later explained in detail in Section \ref{subsec:dynHeteroContri}.
Finally, the solutions from Eq. \ref{eq:forall_Eq} with target matrices of optimized connections (i.e., $minCons$ and $maxCons$) and achievable BWs (i.e., $minBW$ and $maxBW$, having a shape similar to Fig. \ref{subfig:ill3}), are shared with VM workers, where Local Agents consume this information.
\begin{equation} \label{eq:forall_preCondition}
\scalebox{0.7}{$
\begin{aligned}
    & sum_{all} = \sum_{i=1}^N \sum_{j=1}^N DC_{rel_{ij}} - N, \\
    & max_{r_i} = max(DC_{rel_{i}}), \quad \text{for } i = 1, 2, \dots, N.
\end{aligned}
$}
\end{equation}
\begin{equation} \label{eq:forall_Eq}
\scalebox{0.8}{$
\begin{aligned}
    &\quad \quad \quad \quad \quad \quad \quad \forall i,j \in [1, N], \\
    &minCandidate_{ij} = \lfloor DC_{rel_{ij}}/sum_{all} \times (M-1) \rfloor, \\
    &minCons_{ij} = max(minCandidate_{ij}, 1) \times w_{s}, \\
    &maxCons_{ij} = \begin{cases}
    \lceil M \times DC_{rel_{ij}} / max_{r_i} \rceil \times w_{s} & \text{if } i \neq j, \\
    1 & \text{if } i = j,
    \end{cases} \\
    &minBW_{ij} = bw_{ij} \times minCons_{ij} \times r_{vec}, \\
    &maxBW_{ij} = bw_{ij} \times maxCons_{ij} \times r_{vec}.
\end{aligned}
$}
\end{equation}

\subsubsection{Dynamic Local Optimization}\label{subsubsec:localOptim}
The heterogeneous connections from global optimization could become 
sub-optimal due to network dynamics during the query execution. To handle this, local optimization is proposed on Additive Increase and Multiplicative Decrease (AIMD), commonly used in TCP Congestion Control \cite{yang2000general}. We choose AIMD as the initial state of the proposed system begins from maximum throughput and gradually reduces with congestion, thereby reducing the round-trip time (RTT) bias. 
Also, the short invocation interval of the implemented AIMD algorithm makes it apt for variable RTTs in WAN \cite{AIMDFair}. 
Local optimization is run on each VM in each DC. The optimizer first sets the target connections and BWs to maximum values using the optimal range provided by global optimization. 
Then, using lightweight node-level runtime monitoring (e.g., ifTop \cite{iftop}), it goes into either (1) Additive Increase or (2) Multiplicative Decrease mode. 
If monitored BW to a DC is significantly less than the target BW ($\Delta > 100$ Mbps \cite{kim2019delay, chen2021sdtp}), i.e., network congestion, then the target connections and BW are reduced to either the minimum or half of the previous value (multiplicative decrease mode), whichever is higher. If not, the connections to that DC are increased by 1, along with a linear increase of target BW (additive mode), until they reach the maximum configuration provided by the global optimization. Note that this linear increase of BW is similar to the empirically deduced approach of deriving achievable BWs, i.e., using the product of predicted runtime BW and the determined number of connections. In the prior case, if the min-max range from $DC_0$ to other DCs are \{1000, 800, 240\}-\{1000, 1600, 600\} Mbps and \{1, 2, 2\}-\{1, 4, 5\} optimal connections, then decrease mode is set for $DC_0$-$DC_1$ and $DC_0$-$DC_2$ when monitored BW is $<$ 1500 and 500 Mbps, respectively, hinting network congestion. Thereafter, increase mode is set when monitored BW and (current) target BW are similar, hinting improved network availability. Moreover, local optimization checks the data-transfer size of a DC pair as that determines the network utilization. If this size is $<$ 1 MB (derived empirically by monitoring network utilization), then it skips the above checks and avoids the toggle of modes. 
\\ \noindent\textbf{Throttling BW: } To ensure that nearby DCs do not consume bulk of the available network and warrant better network utilization by the parallel connections between distant DCs, the local optimization technique also employs throttling, which limits the maximum achievable BW between nearby DCs. That is, it first computes the threshold (T) for determining BW-rich DCs from a source DC by taking the mean of achievable BWs from that region. Next, for destination DCs with achievable BWs $>$ T, it uses Traffic Control (TC) to limit their achievable BWs to T. Effectiveness of this approach is shown in Section \ref{subsubsec:EvalTC}.

\subsection{Handling Heterogeneity}\label{subsec:dynHeteroContri} 
\subsubsection{Heterogeneous Input Data Distribution}
Skewness is central in GDA because distributed frameworks favor data locality that provides lower latency \cite{Iridium}. Data locality-aware task assignment creates large-scale intermediate data in skewed DCs, demanding higher network capacities in shuffle stages. We propose to resolve this requirement in $2$ phases. First, skewness weights ($w_s$) of input data are collected from underlying storage systems, e.g., HDFS. Second, these weights are used to proportionally re-allocate the optimal range determined by global optimization (Eq. \ref{eq:forall_Eq}). This ensures that higher weightage is given to data-intensive DC regions, reducing network latency in the shuffle stages.

\subsubsection{Heterogeneous Number of DCs} \label{subsubsec:heterodcs}
The model discussed in Section \ref{subsec:wanPredModel},
can predict runtime BWs for different cluster sizes. This is because the training dataset includes different cluster sizes in the range $[2, N_{max}]$, where $N_{max}$ is the maximum cluster size for a GDA application. If in future, $N_{max}$ needs an update, then the RF regressor can be re-trained (on newer cluster sizes) using the warm start.
\subsubsection{Heterogeneous Providers and VMs}\label{subsubsec:heteroVMsDesign}
We introduce \textit{refactoring} and \textit{association} to support GDA applications on multi-cloud providers (e.g., AWS and Azure together) and heterogeneous compute resources (e.g., different numbers of VMs at each DC). Empirical results suggest BWs between such providers and machine types vary proportionally. Thus, to handle this heterogeneity, the determined runtime BWs are adjusted with a refactoring-vector ($r_{vec}$) generated a priori (Eq. \ref{eq:forall_Eq}). Note that refactoring is optional; by default, $r_{vec}$ of all 1s is used. Further, association is used when DC-VM mapping is one-to-many, i.e., BWs are summed to reflect the combined BW of a DC \cite{skyplane}. Once connections are optimized by treating multiple VMs in a DC as $1$ large VM, the global optimization results are proportionally chunked and distributed among workers for local optimization.

\subsubsection{Handling out-of-date Prediction Model}
The prediction error is tracked by intermittently comparing the predicted BWs with actual runtime values. If errors are higher than a pre-configured threshold, a log-based flag is used to signal the need for model retraining. When flagged, GDA applications can then use the additionally collected datasets (as part of model prediction and monitoring processes) to retrain the prediction model using warm start.
\section{WANify Overview}
\begin{figure}[tp]
	\centering
	\includegraphics[width=2.5in]{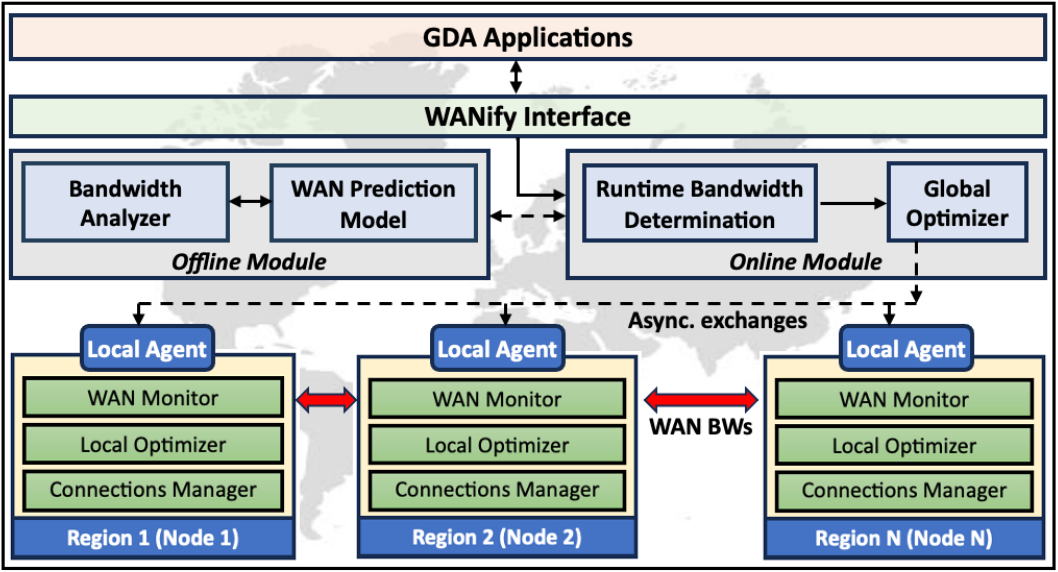}
	\caption{End-to-end architecture of WANify.}
	\label{fig:arch}
\end{figure}
\subsection{WANify Architecture}
Fig. \ref{fig:arch} shows the WANify architecture. 
Any GDA system that transfers data among DCs can reap WANify's benefits 
using the WANify Interface (invoked asynchronously). 

\subsubsection{Offline Module}
The offline module collects BW, trains the prediction model, and predicts 
runtime BW as discussed in Section \ref{subsec:wanPredModel}. 
It consists of $2$ sub-modules.

\noindent\textbf{Bandwidth Analyzer} starts VMs in the configured regions and gathers BW information. It generates datasets to be used for training the WAN Prediction Model. Consequently, it analyzes BWs from several traces collected during the offline monitoring. For both snapshot and runtime BWs, it produces the mapping between DC pairs and their recognized BW so that they can be directly consumed in model training.



\noindent\textbf{WAN Prediction Model} gauges runtime BWs for a given cluster size and for a given combination of snapshot BWs. 
The model is trained on datasets for different cluster sizes,
and thus, it is adaptive to varying cluster sizes in the GDA.

\subsubsection{Online Module}
This module interacts with WANify agents for runtime BW optimization. It has $2$ sub-modules.

\noindent\textbf{Runtime Bandwidth Determination} uses WAN Prediction Model for deriving the runtime BW. Also, it arranges these predicted BWs per DC-pair
for seamless optimizations. 

\noindent\textbf{Global Optimizer} uses a greedy approach to disseminate the capped number of connections per host based on the predicted runtime BWs. That is, it determines an optimal range of connections between the interacting VMs (DCs) for enhancing overall BW, as discussed in Section \ref{subsubsec:globalOptim}.


\subsubsection{Local Agent}\label{subsubsec:localAgent}
Each VM in each DC contains a local agent with $3$ sub-modules. (1) WAN Monitor monitors the network traffic and reports the latest BWs to Local Optimizer. As discussed in Section \ref{subsubsec:localOptim}, (2) Local Optimizer determines the fine-tuned number of optimal connections and target BWs based on the window (or range) received from Global Optimizer and the monitoring metrics reported by WAN Monitor. Lastly, the optimal configurations are fed into (3) Connections Manager, which adds/removes the required connections from the active connection pool.

\section{Evaluation}
\subsection{Experimental Setup}\label{subsec:expSetup}
\noindent\textbf{DCs and VMs Setting:} We use a GDA testbed on AWS with 8 DCs: US East (North Virginia), US West (North California), AP South (Mumbai), AP SE (Singapore), AP SE-2 (Sydney), AP NE (Tokyo), EU West (Ireland), and SA East (São Paulo), shown in Fig. \ref*{fig:gda}. To connect them, we use VPC peering as it provides better performance than the public Internet \cite{skyplane}, which is more realistic for the various GDA workloads. 
For making first-hand comparisons with Kimchi \cite{Kimchi}, we deploy the Spark master on t2.large (2 vCPUs and 8 GB RAM) in US East and workers on t2.medium (2 vCPUs and 4 GB RAM) in each DC.
To minimize CPU bottleneck, we use unlimited CPU bursts
and add \$0.05 per vCPU-hour in the cost values. Note that all query costs include compute, network, and storage costs.
Also, for uniform parallel connections, we use 8 connections per DC-pair as it gives the best minimum and maximum BW between any 2 DCs in our experiments. 

\noindent\textbf{Storage:} 
We use HDFS for GDA input data. The HDFS and Spark master co-reside in US East, while data nodes run on the same worker VMs.
Since AWS S3 is cheaper and more scalable than disk storage,
we created S3 buckets in each region and mounted them as HDFS data nodes, i.e., S3-mounted HDFS.
Note that initial experiments confirmed that using
S3 buckets for input data incur negligible overhead ($< 5\%$) compared to local disks.
Intermediate data is stored and exchanged via each Spark executor's local storage.

\noindent\textbf{Applications:} WANify is tested on various workloads that have been used
in most GDA systems. 
First, to assess the efficacy of different versions of WANify, \textbf{TeraSort} is used. Second, to examine WANify with varying sizes of intermediate data, \textbf{WordCount} is used. Third, to measure the efficacy of WANify against other techniques, \textbf{TPC-DS} is used, consisting of various-sized queries. Lastly, to evaluate the benefits of WANify in WAN-intensive ML convergence, \textbf{Quantization} \cite{quantizationTCC} is used, which adjusts the precision (bits) of floating point for weights and biases based on the available BW without compromising model accuracy.

\noindent \textbf{Input Size:} WANify is tested on 100 GB of input data for TeraSort and TPC-DS. Note that we also use 40 GB of TPC-DS input to make seamless comparisons with Kimchi \cite{Kimchi}. For ML-based tests, MNIST dataset is used with union transformations in PySpark to generate a large input of $\sim6.8$ GB. Further, to control the intermediate data size for WordCount, we use inputs in the range [100-600] MB with all distinct words. Lastly, to test WordCount on uniform and skewed inputs, we use HDFS block size of 64 MB.
\\ \noindent\textbf{Baselines:} 
We compare WANify's approaches against other simple approaches used in existing GDA systems, such as
1) statically and independently measured BW, 
2) a single connection for data transfer, and
3) a uniform number of parallel connections. 
Further, we evaluate the WANify prototype
using $3$ GDA systems, Tetrium \cite{Tetrium}, Kimchi \cite{Kimchi}, and SAGQ \cite{quantizationTCC}; SAGQ being an ML-inference application in GDA.
Note that SAGQ is implemented on Spark using PySpark and modified elephas libraries. 
All results are a mean of 5 runs, plotted with standard errors as error bars.
\\ \noindent\textbf{WAN BW Prediction Model:} 
To train the prediction model, we first run Bandwidth Analyzer at different times over a week. For various cluster sizes, we collect 600 datasets, each having (1) short-duration snapshot BWs (along with other features listed in Table \ref{tab:predictFeatures}) and (2) dynamically measured BWs. The collected datasets show an overall standard deviation of $\sim184$ Mbps for the dynamically measured BWs, which is expected with the heterogeneous DCs and large dynamics in WAN \cite{FluctuatingBWsIMC}. These datasets are used to train the RF component of WAN Prediction Model. We explore several configurations of estimators in the RF and eventually set it to 100 as this yields the best training accuracy of 98.51\% (derived from historical training metrics). Later in Section \ref{subsec:validation} and \ref{handlingHetero}, we assess the accuracy of the prediction model. Note that all features in Table \ref{tab:predictFeatures} were significant during model training. Also, it cost us around $\$150$ for dataset collection and model training on AWS.

\subsection{Gauging Runtime BW}\label{subsec:singleConImpr}
To evaluate the benefits of snapshot-assisted runtime BW prediction in GDA, 
we use $3$ classes of TPC-DS queries: light-weight: query 82 \cite{Smartpick}, average-weight: queries 11 and 95 \cite{Kimchi, Smartpick}, and heavy-weight: query 78 \cite{tpcdsWorkloadAnalysis}. 
We first measure single-connection static-independent BWs using iPerf and feed them into 
\textit{unmodified} Tetrium and Kimchi (for default task placement) as a baseline.
Next, for comparison, we feed single-connection static-simultaneous BWs
and predicted single-connection runtime BWs into them. 

Table \ref{tab:singleConnRealtimeBW} shows the performance improvement (Perf.) and cost reductions (Cost) for all queries. 
The results clearly show that existing GDA systems would
improve performance with reduced cost if they consider simultaneous network BWs while
making decisions. That is, query latencies are reduced by up to $\sim16\%$ and $\sim18\%$ using simultaneously 
measured BWs and predicted BWs, respectively.
Such performance gains occur because of better minimum BW during query execution ($\sim1.5 \times$ improvement for average-weight and heavy-weight queries).
Further, the results show cost reduction by up to $\sim5.2\%$, thanks 
to query performance improvement for both cases, i.e., reduced compute resources cost.
Note that the differences in network cost are negligible as data transferred over WAN remains almost the same.
More importantly, the results confirm that predicted runtime BWs yield similar benefits to static-simultaneous BWs, 
which shows the efficacy of WANify's prediction model. 
In terms of monitoring costs for these queries, the prediction model requires only $\sim\$5$ compared to $\sim\$80$ for obtaining static-simultaneous BWs ($\sim94\%$ savings). Note that we compare predicted and static-simultaneous BWs in greater detail in Section \ref{handlingHetero}.

These results show that WANify helps GDA systems make better decisions by gauging runtime BW precisely using snapshots, which reduces query latency and cost while notably saving BW monitoring costs.
Note that such gains use a single connection, and thus, they can be enhanced further by using parallel connections (Section \ref{subsec:multiConImpr}).
\begin{table*}
    \quad \quad \quad
    \begin{minipage}[b]{0.55\linewidth}
	\centering
	\caption{Performance-cost improvements against static BWs} 
	\scalebox{0.65}{\begin{tabular}{|c|c|c|c|c|c|c|c|c|}
			\hline
			& \multicolumn{4}{c|}{\begin{tabular}{c}\textbf{Tetrium}\end{tabular}} & \multicolumn{4}{c|}{\begin{tabular}{c}\textbf{Kimchi}\end{tabular}} \\
			\cline{2-9}
                & \multicolumn{2}{c|}{\begin{tabular}{c}\textbf{Static-simultaneous}\end{tabular}} & \multicolumn{2}{c|}{\begin{tabular}{c}\textbf{Predicted (WANify)}\end{tabular}} & \multicolumn{2}{c|}{\begin{tabular}{c}\textbf{Static-simultaneous}\end{tabular}} & \multicolumn{2}{c|}{\begin{tabular}{c}\textbf{Predicted (WANify)}\end{tabular}} \\
			\hline
   
			\multirow{-4}{*}{\begin{tabular}{c}\textbf{Query \#}\end{tabular}} & Perf. (\%) & Cost (\%) & Perf. (\%) & Cost (\%) &
            Perf. (\%) & Cost (\%) & Perf. (\%) & Cost (\%) \\
			\hline
			82 & $\sim$1 & 3.9 & $\sim$1 & 3.9 & $\sim$1 & 5.2 & $\sim$1 & 5.2 \\
			\hline
			95 & 9.5 & 2.4 & 8 & 2 & 10.2 & 2.1 & 11.7 & 2.8 \\
			\hline
			11 & 11 & 3.7 & 10.2 & 3.5 & 16 & 3 & 18 & 3.7 \\
			\hline
			78 & 12.5 & 3 & 14 & 3.1 & 14 & 1.4 & 13 & 1.1 \\
			\hline
	\end{tabular}}
	\label{tab:singleConnRealtimeBW}
    \end{minipage}\quad \quad \quad \quad \quad 
\begin{minipage}[b]{0.25\linewidth}
	\centering
	\includegraphics[height=0.66in]{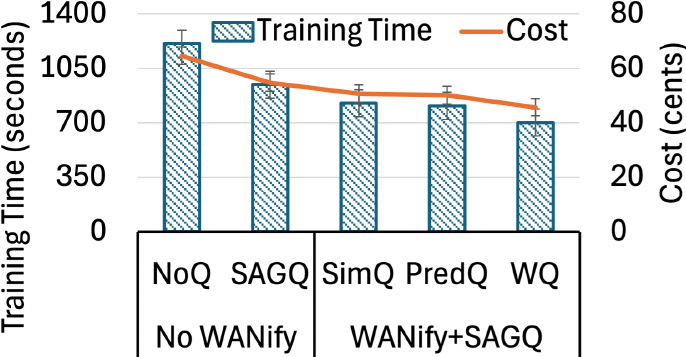}
	\captionof{figure}{Impact on ML in GDA.}
	\label{fig:deepLearning}
\end{minipage}
\end{table*}


\subsection{Parallel Data Transfer (PDT)}
This section evaluates the gains from using parallel data transfer alone and thus, avoids WAN-aware GDA systems.
\subsubsection{Comparing Data Transfer Approaches} \label{subsubsec:EvalTC}
We use TeraSort to compare the single connection \textit{vanilla-Spark} on locality-aware scheduling (i.e., No WAN-aware) against 3 approaches on predicted runtime BWs: (1) \textbf{WANify-P}: uniform parallel connections, (2) \textbf{WANify-Dynamic}: heterogeneous parallel connections, and (3) \textbf{WANify-TC}: heterogeneous parallel connections and dynamic BW throttling using TC for BW-rich links (as discussed in Section \ref{subsubsec:localOptim}).

\begin{figure}[tp]
	\centering
	\subfigure[Latency/cost metrics]{\label{subfig:teraLatCost}\includegraphics[width=1.32in]{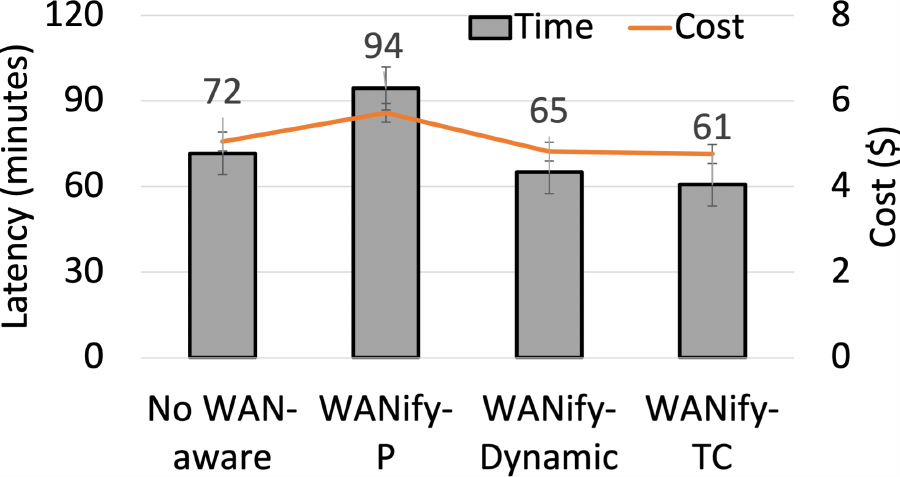}}
	\quad
	\subfigure[Min. BW metrics]{\label{subfig:teraBW}\includegraphics[width=1.32in]{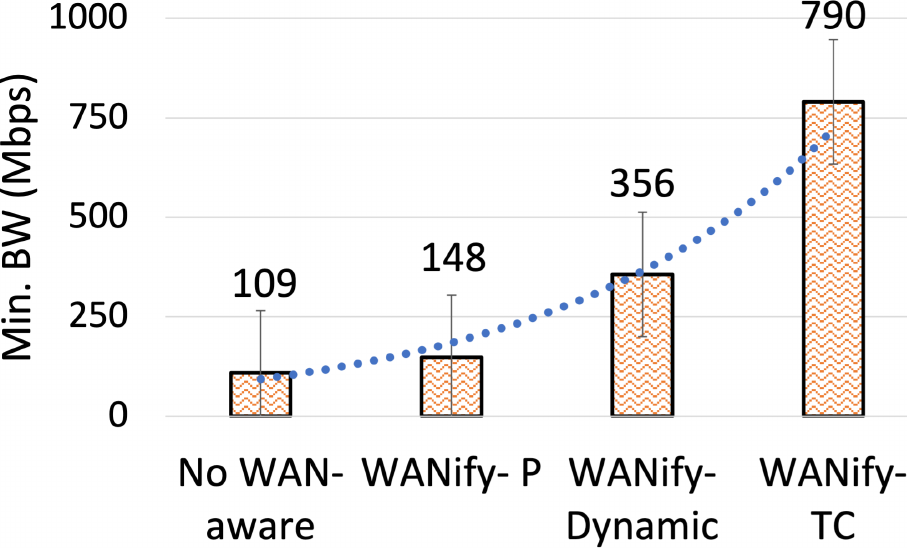}}
	\caption{Comparing different parallel approaches.}
	\label{fig:terasort}
    \quad
	\subfigure[Latency/cost metrics]{\label{subfig:wrdCntlat}\includegraphics[width=1.32in]{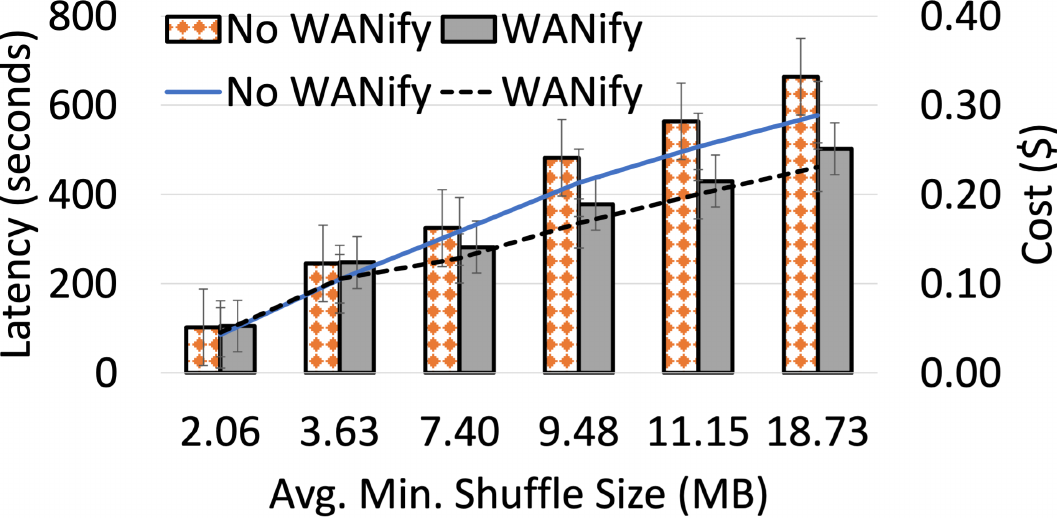}}
	\quad
	\subfigure[Min. BW metrics]{\label{subfig:wrdCntBW}\includegraphics[width=1.32in]{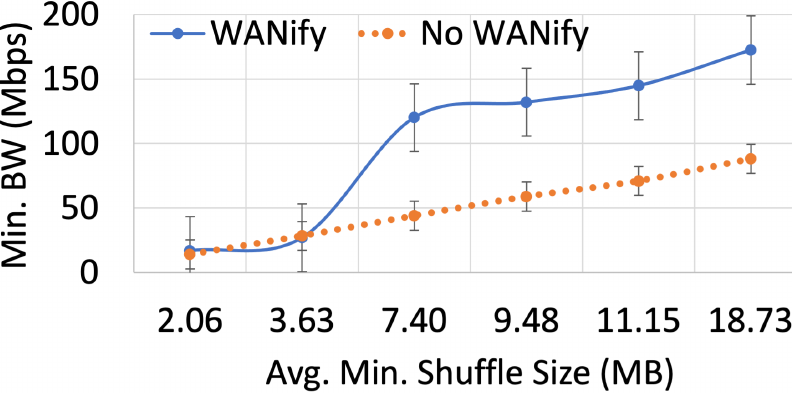}}
	\caption{Efficacy against various shuffle sizes.}
	\label{fig:wordCnt}
\end{figure}

Fig. \ref{fig:terasort} shows the latency, cost, and minimum BW for each approach. WANify-P has increased latency and cost with no key improvements to the minimum BW
due to network congestion. WANify-Dynamic beats WANify-P with optimal heterogeneous connections, increasing the minimum BW to 356 Mbps. Besides, the default model, i.e., WANify-TC, gives the best performance (61 minutes), cost (\$4.7), and minimum BW (790 Mbps) by dynamically throttling BW between BW-rich DCs. Thus, dynamic BW throttling in local agents yields the optimal network and query performance.
\subsubsection{Comparing Different Intermediate Data Sizes} \label{subsubsec:diffinter}
We use WordCount to examine how intermediate data sizes affect the performance of parallel data transfer in WANify, i.e., WANify-TC, by comparing the results with vanilla-Spark, which uses a single connection. 
To control intermediate data between each pair of DCs, we write 
a Python script to generate all distinct words for a given size. 

Fig. \ref{subfig:wrdCntlat} shows the latency and cost decrease for WANify as intermediate data increases from 3.63 to 7.4 MB and beyond. Fig. \ref{subfig:wrdCntBW} shows the change in minimum BW for respective data sizes. For lower data sizes, i.e., 2.06 and 3.63 MB, both WANify and No WANify show similar latency, cost, and minimum BW since the required WAN capacity is low. For data sizes $>$ 7.4 MB, WANify reduces latency and cost with improved minimum cluster BW (120 $\sim$ 172 Mbps) by using heterogeneous connections. Thus, WANify effectively handles various sizes of intermediate data.

\subsection{Benefits from Predicted BWs and PDT}\label{subsec:multiConImpr}
\begin{figure}
        \centering
	\subfigure[Query latencies]{\label{subfig:tpcdsMultiConLat}\includegraphics[width=1.36in]{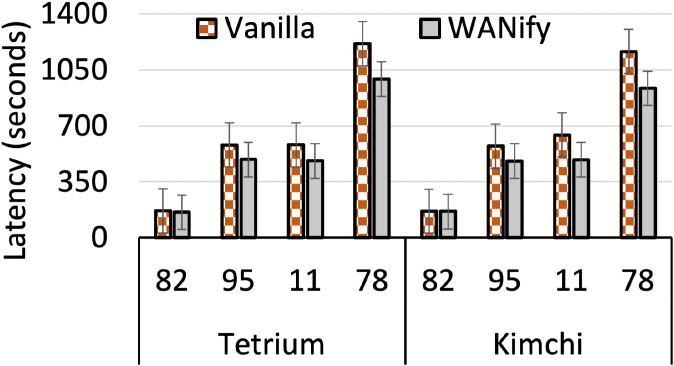}}
	\quad
	\subfigure[Query costs]{\label{subfig:tpcdsMultiConCost}\includegraphics[width=1.36in]{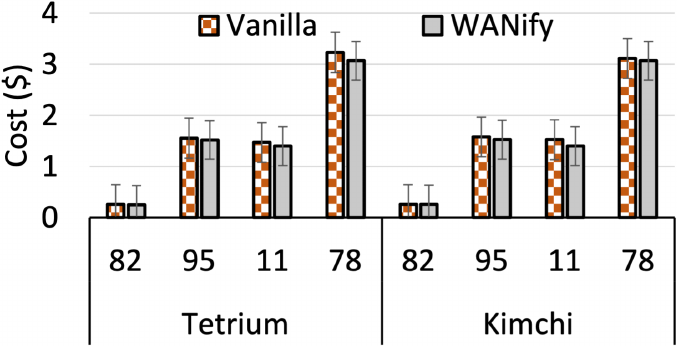}}
	\caption{Comparing state-of-the-art systems on TPC-DS.}
	\label{fig:tpcdsMultiCon}
\end{figure}

Here, we evaluate how WANify helps existing GDA systems, i.e., Tetrium and Kimchi, to further lower query latency and cost by utilizing predicted runtime BWs and heterogeneous parallel connections. 
Fig. \ref {subfig:tpcdsMultiConLat} and Fig. \ref{subfig:tpcdsMultiConCost} show the latencies and cost, respectively, for the TPC-DS queries, 82, 95, 11, and 78, with/without WANify. 
For queries 95, 11, and 78, the results show that WANify helps
GDA systems reduce latencies (by up to $24\%$) and costs (by up to $8\%$) 
compared to baselines, i.e., no WANify. 
Note that cost reduction arises from the reduction in query latency, yielding reduced compute costs 
but not network costs as discussed in Section \ref{subsec:singleConImpr}.
Lastly, we observed a $3.3 \times$ increase in the minimum BW 
when WANify is used in GDA systems.
Although we got similar results from 40 GB of TPC-DS input data (originally used in Kimchi \cite{Kimchi}), we only report findings from the larger 100 GB dataset due to space constraints. Overall, the results confirmed that WANify helps boost the achievable BW for GDA systems by employing optimal number of heterogeneous parallel connections between the interacting DCs, improving performance with reduced cost. 

\subsection{Validation of WANify's Design}\label{subsec:validation}
\begin{figure}[tp]
	\centering
	\subfigure[Ablation study]{\label{subfig:ablation}\includegraphics[height=0.9in]{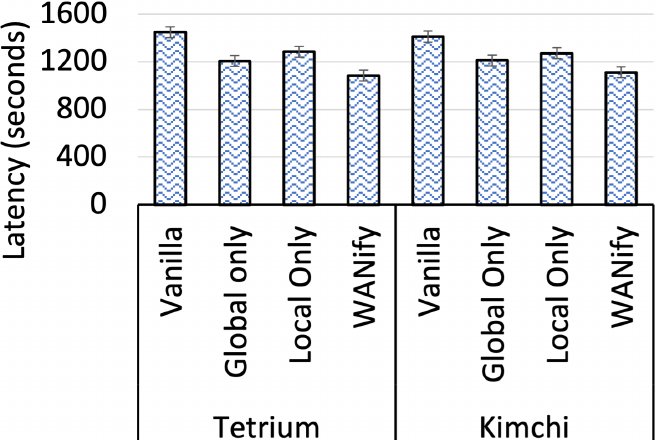}}
	\quad
	\subfigure[Impact of prediction error]{\label{subfig:impactError}\includegraphics[height=0.9in]{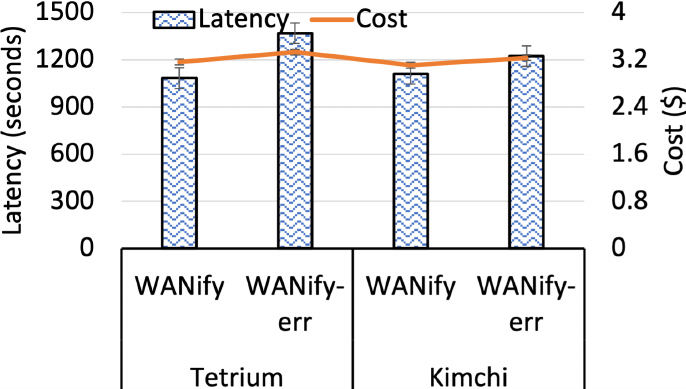}}
	\caption{Efficacy of WANify's components.
	\label{fig:roleComps}}
\end{figure}

To test the gains from global and local optimization, we run an ablation study for query 78 on $4$ WANify variants: (1) \textbf{Vanilla}: with unmodified GDA system, (2) \textbf{Global only} where only global optimization is enabled, (3) \textbf{Local only} where only local optimization is enabled with a static min-max range of $1-8$ connections for each DC-pair, and (4) \textbf{WANify} where all optimizations are enabled. Fig. \ref{subfig:ablation} shows the results for Tetrium and Kimchi. Note that the values listed here differ in magnitude but exhibit a similar pattern to those in Section \ref{subsec:multiConImpr}, as they were re-run to examine WANify's efficacy at different times of the day. The Global only approach provides $\sim16\%$ better latency than Vanilla systems, since the optimal heterogeneous connections yield $\sim1.2\times$ better minimum BW. The Local only approach provides $\sim11\%$ better latency and $\sim1.1\times$ better minimum BW than Vanilla systems because it still heterogeneously fine-tunes the connections at runtime. However, it offers $\sim5\%$ worse latency than the Global only approach as it is unaware of inferred DC closeness, causing sporadic unfair network utilization by nearby DCs (Section \ref{subsec:optimalconnect}). Lastly, WANify-enabled GDA system achieves the best reduction in latency ($\sim 23\%$) than Vanilla systems, which shows the importance of combined global and local optimization.

To assess the impact of BW prediction errors, we randomly add/subtract significant BW values (100 Mbps \cite{kim2019delay, chen2021sdtp}) to/from the WANify predicted BWs, i.e., WANify-err, and compare its behavior with the WANify system. Fig. \ref{subfig:impactError} shows the results for query 78, where WANify-err increases latency and cost by $\sim 18\%$ and $\sim 5\%$, respectively. Also, WANify-err decreases the minimum cluster BW by $\sim 38\%$, which explains the observed performance-cost bottlenecks. 
These results show that WANify's real-time snapshot-enabled prediction model is highly accurate.

The above findings affirm the soundness of WANify's design, which achieves better cost-performance gains using runtime BWs and optimal heterogeneous connections.
\subsection{Benefits from WANify in ML Applications}\label{subsec:evalGeoDML}
This test compares WANify's gain against various WAN-aware ML models with quantization \cite{quantizationTCC}. We consider the MNIST dataset on $4$ deep learning models: (1) \textbf{NoQ}: No quantization, (2) \textbf{SAGQ}: Static independent BWs driven quantization, (3) \textbf{SimQ}: Simultaneous BWs driven quantization, and (4) \textbf{PredQ}: Predicted BWs driven quantization.
We compare them against WANify-enabled quantization (\textbf{WQ}) with parallel heterogeneous connections. Each model consists of 3 Dense layers, 3 Activation layers, and 2 Dropout layers, and is trained on the 8-DC Spark cluster. 

Fig. \ref{fig:deepLearning} shows the observed training overhead and overall costs for these models when each runs for $10$ epochs with a test accuracy of $\sim97\%$. Note that the cost reduction results from lower computation time, since network cost remains the same for all the models ($\sim20$ cents). It is evident that compared to NoQ, SAGQ reduces training time and cost by $\sim22\%$ and $\sim15\%$, respectively, as shown in the previous work \cite{quantizationTCC}.
More importantly, compared to SAGQ, SimQ and PreQ further reduce training time and cost by $13\sim14.5\%$ and $7\sim8\%$, respectively, which shows the importance of accurate simultaneous BWs. Lastly, WQ provides the best performance ($\sim26\%$ and $13\%$) and cost ($16\%$ and $9\%$) improvements compared to SAGQ and PredQ, respectively by leveraging a $2\times$ boost to the minimum BW.

These results affirm WANify's high accuracy prediction (with alike SimQ and PredQ gains) and echo the benefits of the heterogeneous approach. Also, they show WANify's generalizability to diverse GDA workloads, including ML.
\begin{figure}[tp]
        \centering
	\subfigure[Accuracy of local optimizer]{\label{subfig:accWithoutErr}\includegraphics[width=1.4in]{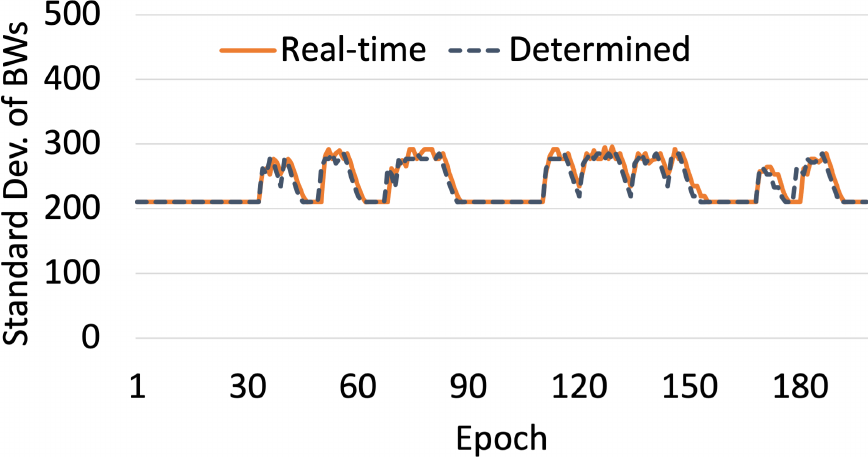}}
	\quad
	\subfigure[Accuracy with 20\% error]{\label{subfig:accWithErr}\includegraphics[width=1.4in]{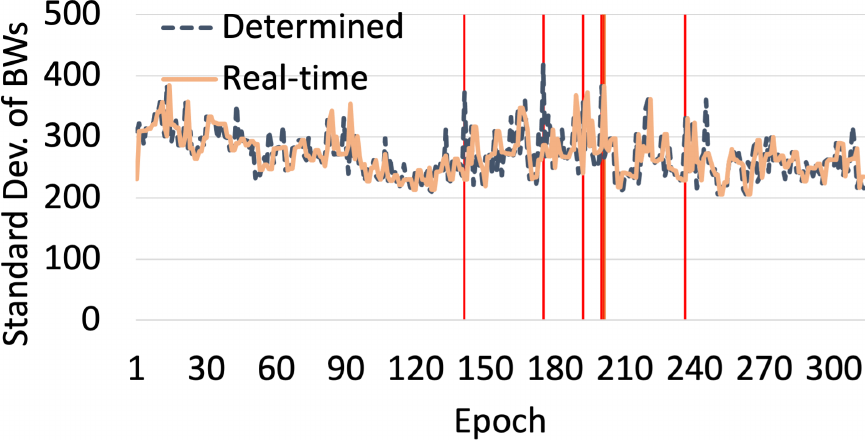}}
	\caption{SD of BWs. Verticals show $\mathbf{> 100}$ Mbps deltas.} 
	\label{fig:distOptAcc}
    \end{figure}
    \begin{figure}[tp]
    \centering
	\subfigure[Tetrium improvements]{\label{subfig:tetriumSkewed}\includegraphics[width=1.3in]{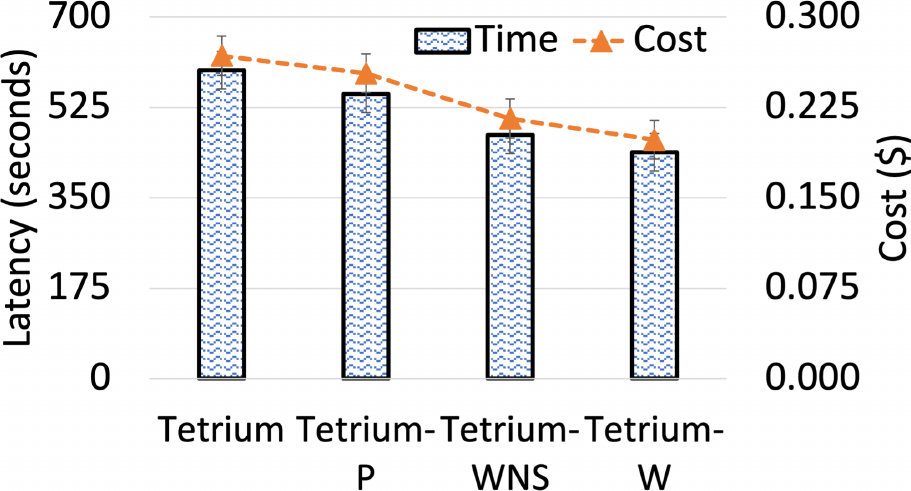}}
	\quad
	\subfigure[Kimchi improvements]{\label{subfig:kimchiSkewed}\includegraphics[width=1.3in]{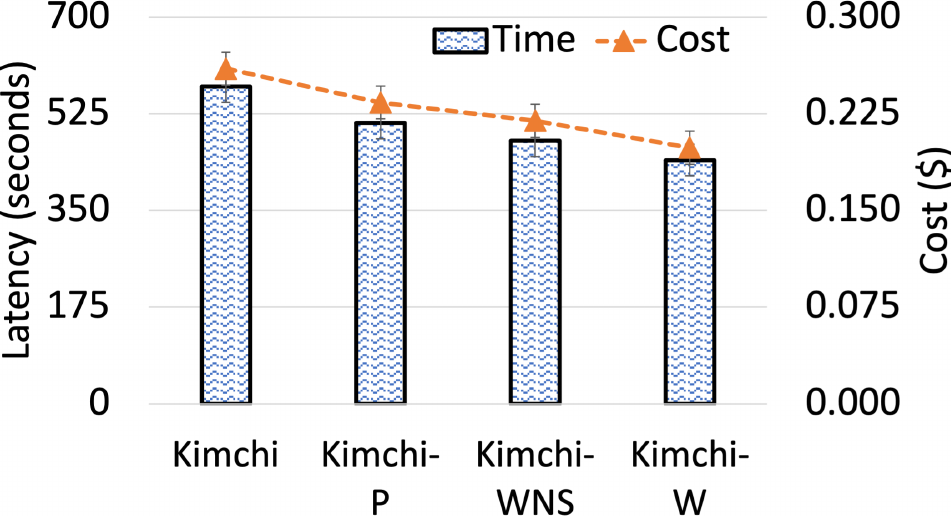}}
	\caption{Efficacy of WANify in handling skewed inputs.}
	\label{fig:skewed}
\end{figure}
\subsection{Handling Dynamics}\label{subsec:handlingDyn} 
To evaluate WANify in the presence of dynamics, we use WANify-enabled Tetrium and compute standard deviation (SD) of WANify-determined local-optimizer BWs from one region (US East) to every other region. Next, we compare it with the SD of actual runtime BWs monitored using ifTop. 

Fig. \ref{subfig:accWithoutErr} shows the variation of these SDs across several epochs, where an epoch refers to the 5-second interval at which the local optimizer updates the target BWs based on the AIMD technique (Section \ref{subsubsec:localOptim}). It is observed that WANify accurately models the runtime BWs by decreasing the target BWs when congestion is detected and increasing the target BWs otherwise. Note that this increase and decrease of target BWs is based on the individual DC-pair measurements and thus the SD of all link-BWs represents the direction in which the current network state is moving. 

For comparison, we add 20\% random errors to the derived optimal connections and target BWs and record instances where the change from actual runtime values is \textit{significant}, i.e., $> 100$ Mbps \cite{kim2019delay, chen2021sdtp}. 
Fig. \ref{subfig:accWithErr} shows $6$ significant (marked with verticals)
and other minor differences across the epochs. Moreover, the number of epochs increase as the model needs to frequently update the target BWs with random errors since they no longer align well with the runtime values.
Note that the results in Section \ref{subsec:multiConImpr} include handling of dynamics for best performance. 
\subsection{Handling Heterogeneity}\label{handlingHetero}

\subsubsection{Heterogeneous Data Distribution}
To evaluate skewed data performance for WANify-enabled systems, we run WordCount with 600 MB input by moving HDFS blocks from other DCs to US East, US West, AP South, and AP SE. 
Next, we compare $4$ approaches on predicted runtime BWs:
(1) \textbf{Tetrium}: single connection, (2) \textbf{Tetrium-P}: uniform parallel connections, (3) \textbf{Tetrium-WNS}: WANify-enabled Tetrium without factoring skewness, and (4) \textbf{Tetrium-W}: WANify-enabled Tetrium with consideration of skewness.

Fig. \ref{subfig:tetriumSkewed} shows cost-performance comparisons for each approach.  
Average latency for Tetrium-W improves by 26.5\%, 20.3\%, and 7.1\%, and average cost reduces by 26\%, 21.7\%, and 8.1\%, compared to Tetrium, Tetrium-P, and Tetrium-WNS, respectively. Also, Tetrium-W has 1.2 - 2.1 $\times$ higher minimum BW than other approaches. Fig. \ref{subfig:kimchiSkewed} shows similar results with Kimchi. 
These results show that WANify considers data distribution while predicting 
BW and making decisions, which reduces query latency and cost. 
\subsubsection{Heterogeneous Number of DCs}
\label{subsubsec:heterodcs}
Here, we test the model's accuracy with heterogeneous multi-DC clusters on AWS. For variable DCs in the cluster and with $1$ VM at each DC, the pair-wise BWs are noted using: (1) static-independent measurements, (2) WANify predicted BWs, and (3) runtime BWs. 
Each BW in (1) and (2) is compared with (3) to record 
significant differences ($>$ 100 Mbps \cite{kim2019delay, chen2021sdtp}).

\begin{figure}[tp]
	\centering
	\subfigure[Heterogeneous DCs]{\label{subfig:multiDCModelAcc_AWS_1}\includegraphics[height=0.65in]{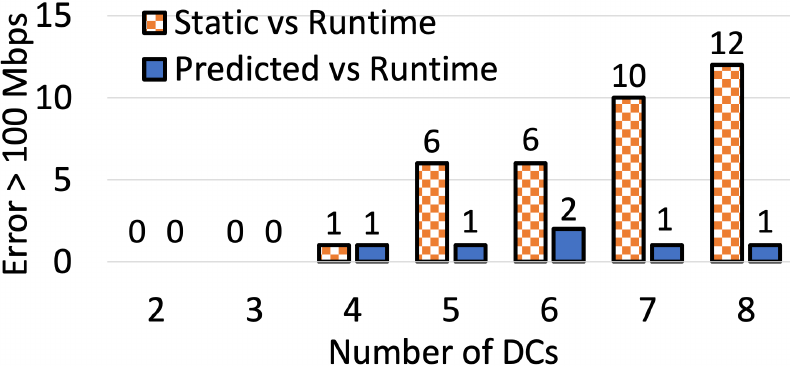}}
	\quad
	\subfigure[Heterogeneous VMs]{\label{subfig:multiDCmVMModelAcc_AWS}\includegraphics[height=0.705in]{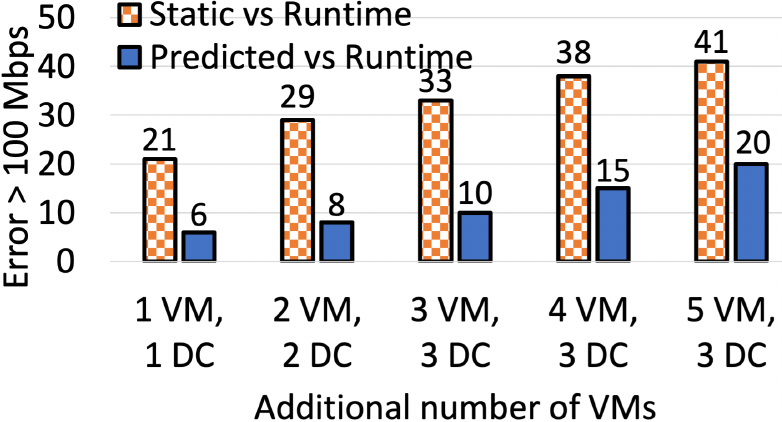}}
	\caption{Comparing accuracy of predicted runtime BWs.
	\label{fig:multiDCModelAcc_AWS}}
\end{figure}

Fig. \ref{subfig:multiDCModelAcc_AWS_1} shows that WANify's predicted BWs beat the static-independent BWs for various cluster sizes. 
Thus, lower prediction errors would improve cost-performance gains for various cluster sizes, as shown in Section \ref{subsec:singleConImpr}. Moreover, the results agree with our system model (Section \ref{subsec:illEx}), which aims to reduce the significant BW differences from the runtime scenario. Lastly, these results align well with the WAN traffic analysis \cite{FluctuatingBWsIMC} that shows predictability on the scale of minutes. Note that RF can accurately predict runtime BWs for a highly dynamic WAN in GDA \cite{Gaia, Tetrium, FluctuatingBWsIMC} because it uses real-time snapshots and employs bias-variance tradeoff in ensemble learning, aiding in generalization \cite{belkin2019reconciling}. 

\subsubsection{Heterogeneous Number of VMs}
To examine prediction accuracy with heterogeneous compute resources, $1\sim5$ additional VMs are assigned to 3 randomly selected AWS DCs, i.e., non-uniform VM deployment. We compare actual runtime BW with (1) statically measured BW and (2) predicted BW as done in Section \ref{subsubsec:heterodcs}. 

\noindent\textbf{Accuracy:} Fig. \ref{subfig:multiDCmVMModelAcc_AWS} shows that predicted BW has a higher accuracy than statically measured BW, with a lesser number of significant BW differences from the actual runtime values.
This confirms that predicted runtime BW is significantly closer 
to actual runtime BW than static BW. 
Similar tests were run in a multi-cloud environment, 
AWS and Google Cloud Platform (GCP) together, using similar VMs, i.e., 
AWS t2.medium and GCP e2-medium. 
While we observed similar results, we omit them here due to space constraints.

Although WANify beats the static BW approach, it is not 100\% accurate like most prediction systems because of network dynamics in GDA \cite{Gaia}. Yet, as evidenced by the results and as discussed in Section \ref{subsubsec:localAgent}, WANify's local agents account for this marginal error.

\noindent\textbf{Benefits in GDA:} 
We run TPC-DS query 78 with heterogeneous compute capacities on the 8 AWS DCs, 
where an extra t2.medium VM is added in US East. 
Since Tetrium works with heterogeneous compute,
we compare Tetrium on predicted single connection runtime BWs
(Tetrium-r) against vanilla-Tetrium with static and independent BWs. Tetrium-r results in $5\%$ lower latency and $1\%$ lower cost than vanilla-Tetrium with a $1.2 \times$ speed-up in the minimum BW.
This shows that WANify accurately models heterogeneous VMs 
in GDA, as discussed in Section \ref{subsubsec:heteroVMsDesign}. Also, WANify-enabled Tetrium with both predicted runtime BWs and parallel heterogeneous connections, achieves the best benefits with $15\%$ lower latency, $7.4\%$ lower cost, and $2 \times$ speed-up in the minimum BW compared to vanilla-Tetrium.
\section{Conclusion}
In this paper, we proposed WANify, a runtime WAN enhancement framework for GDA. 
WANify uses real-time snapshots to gauge and predict accurate runtime WAN BW that reduces query latency and cost while notably saving monitoring costs. Moreover, it determines optimal heterogeneous parallel connections, allowing GDA systems to further lower query latency and cost by fully exploiting available BW.
To avoid bottlenecks from network dynamics, e.g., network congestion,
WANify uses AIMD technique that dynamically adjusts the active number of connections at each DC region.
Also, WANify adaptively tunes the determined heterogeneous network connections to handle heterogeneity in GDA such as 
varying cluster sizes, non-uniform compute capacities, 
and input data skewness.
Experimental results on AWS show the precision of WANify in 
using real-time snapshots and predicting the runtime WAN BW with an accuracy of 98.51\%. 
The results also confirm that WANify enables existing 
GDA systems (on diverse workloads) to reduce latency (up to 26\%) and cost (up to 16\%)
by enhancing the weak WAN links while handling network dynamics and heterogeneity efficiently. 



\balance
\IEEEtriggeratref{35}
\bibliographystyle{IEEEtranS}
\bibliography{reference}

\end{document}